# Evolution of ferroelectricity with annealing temperature and thickness in sputter deposited undoped HfO$_2$ on silicon

Md Hanif Ali, Adityanarayan Pandey, Rowtu Srinu, Paritosh Meihar, Shubham Patil, Sandip Lashkare, Udayan Ganguly, *Senior Member, IEEE*

*Abstract*— Ferroelectricity in sputtered undoped-HfO$_2$ is attractive for composition control for low power and non-volatile memory and logic applications. Unlike doped HfO$_2$, evolution of ferroelectricity with annealing and film thickness effect in sputter deposited undoped HfO$_2$ on Si is not yet reported. In present study, we have demonstrated the impact of post metallization annealing temperature and film thickness on ferroelectric properties in dopant-free sputtered HfO$_2$ on Si-substrate. A rich correlation of polarization with phase, lattice constant, and crystallite size and interface reaction is observed. First, anneal temperature shows o-phase saturation beyond 600 °C followed by interface reaction beyond 700 °C to show an optimal temperature window on 600-700 °C. Second, thickness study at the optimal temperature window shows an alluring o-phase crystallite scaling with thickness till a critical thickness of 20 nm indicating that the films are completely o-phase. However, the lattice constants (volume) are high in the 15-20 nm thickness range which correlates with the enhanced value of 2P$_r$. Beyond 20 nm, crystallite scaling with thickness saturates with the correlated appearance of m-phase and reduction in 2P$_r$. The optimal thickness-temperature window range of 15-20 nm films annealed at 600-700 °C show 2P$_r$ of ~35.5 µC/cm$^2$ is comparable to state-of-the-art. The robust wakeup-free endurance of ~10$^8$ cycles showcased in the promising temperature-thickness window identified systematically for non-volatile memory applications.

## I. INTRODUCTION

The excellent scalability, Complementary Metal Oxide Semiconductor (CMOS) compatibility, non-volatility, low-power and high speed operation of HfO$_2$-based ferroelectric devices, viz., ferroelectric random access memory (FERAM), ferroelectric field effect transistor (FeFET), and ferroelectric tunnel junction (FTJ) make it more suitable for neuromorphic applications [1].

Since the discovery of ferroelectricity in thin atomic layer deposited (ALD) Si-doped HfO$_2$ films by Böscke *et al*. in 2011 [2], many other dopants such as Zr, La, Y, Gd, Ce, Sr, and Al, etc. are experimentally demonstrated to enhance ferroelectricity in HfO$_2$ [3]–[6]. It is established that the non-centrosymmetric polar-orthorhombic phase (o-phase, *Pca*2$_1$) in doped-HfO$_2$ is responsible for ferroelectricity [7]. In addition to doping, the surface/interface and grain boundary energy, strain, and the oxygen vacancies present in the film also enhance the stabilization of the o-phase and different other non-polar polymorphs of HfO$_2$ such as tetragonal (t-phase, *P4$_2$/nmc*), non-polar orthorhombic (O$_{II}$-phase, Pbca), monoclinic (m-phase *P2$_1$/c*) and cubic (c-phase, Fm$\bar{3}$m) are also present in nature [8]–[10]. Several groups reported that deposition temperature (T$_d$), annealing condition and film thickness can tune the grain size which could form o-phase in doped FE-HfO$_2$ film [11]–[13]. However, Polakowski *et al*. [14] and Kim *et al*. [15] have shown that the low T$_d$ and film thickness (<10 nm) decrease the grain size and stabilize the ferroelectric o-phase in pure HfO$_2$.

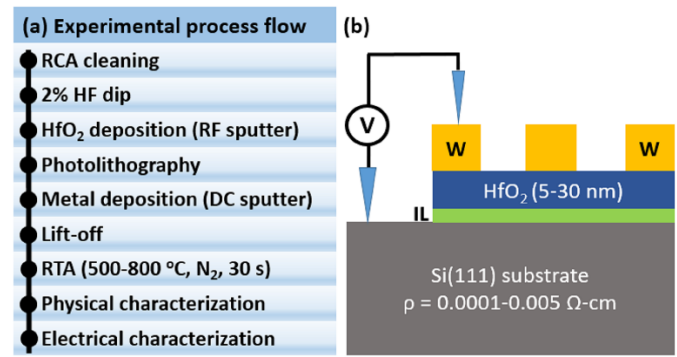

**Figure 1: (a) Experimental process flow, (b) Device schematic**

Several deposition techniques such as ALD, CSD (Chemical solution deposition), and physical vapor deposition (Sputter and Pulsed laser deposition (PLD)) are used to deposit ferroelectric HfO$_2$-based thin films and among them, ALD based doped ferroelectric HfO$_2$ thin films are well optimized [16]. In comparison to others, control over deposition ambient, power, and temperature makes sputter as a better deposition technique with the advantage of low carbon contamination which is a major

This work is supported in part by DST Nano Mission, Ministry of Electronics and Information Technology (MeitY), and Department of Electronics, through the Nano-electronics Network for Research and Applications (NNETRA) project of Govt. of India. It was performed at IIT Bombay Nanofabrication Facility.

All the authors are with the Department of Electrical Engineering, Indian Institute of Technology Bombay, Mumbai, 400076, India (e-mail: hanif.vlsi@gmail.com, anbp.phy@gmail.com, udayan@ee.iitb.ac.in).

challenge in ALD and CSD [17]. The precise doping in HfO$_2$ is also very difficult to control and the doping window to get good ferroelectricity is very narrow within 10% for dopants other than Zirconia for which it is ~ 50% [18]. Thus, undoped FE-HfO$_2$ avoids the concerns of dopant control. Mittmann *et. al.* [19] reported polarization of 2P$_r$ ~ 20 µC/cm$^2$ in an undoped sputtered HfO$_2$ film (20 nm, TiN/HfO$_2$/TiN stack) annealed at 800 °C. Recently, Gronenberg *et. al.* [20] demonstrated the annealing effect on a fixed thickness (~9.8 nm) of undoped sputtered FE-HfO$_2$, where they reported wake-up free remnant polarization (P$_r$) below 10 µC/cm$^2$ to above 20 µC/cm$^2$ with strong wakeup effect. However, study related to the effect of both annealing and film thickness on ferroelectricity in undoped HfO$_2$ deposited on Si by sputter is scarcely reported in literature.

In this work, we have demonstrated a temperature-thickness window with high 2P$_r$ of ~35.5 µC/cm$^2$ in sputter deposited undoped-HfO$_2$ film annealed in N$_2$ ambient. In this window, temperature and thickness ranges are 600-700 °C and 15-20 nm, respectively. The endurance measurement shows the wakeup-free endurance up to ~10$^8$ cycles with a significant 2P$_r$ window of ~17 µC/cm$^2$. Finally, our results are compared with previously reported results on undoped FE-HfO$_2$ which (Table-II) reveals that this work shows better ferroelectricity with comparatively low processing temperature.

## II. EXPERIMENTAL SECTION

RCA cleaned heavily doped (resistivity ~ 0.0001 – 0.005 Ω-cm) p-type Si(111) wafers are dipped in to 2% HF diluted solution to remove native oxide. Further, it is immediately subjected to the sputter and ~15 nm HfO$_2$ film (99.9% pure HfO$_2$ target) is deposited on Si substrate with 50-watt RF power and 3 mTorr chamber pressure at room temperature (RT) using AJA Sputter ATC-2200. Then, photolithography process is carried out to pattern the devices (50 µm x 50 µm) followed by ~50 nm Tungsten (W) as top electrode (TE) deposited using DC sputter. Subsequently, samples are annealed at 500-800 °C in N$_2$ ambient for 30 s by rapid thermal annealing process. Further, undoped HfO$_2$ film of varying thicknesses (5, 10, 15, 20 and 29 nm) are fabricated and then annealed at 600 and 700 °C in N$_2$ ambient for 30 s. After annealing, an interlayer of SiO$_x$ is formed at the HfO$_2$-Si interface in all samples and makes the stack metal-ferroelectric-insulator-semiconductor (MFIS) which is evaluated later. The experimental process flow and the device schematic are shown in fig. 1. The grazing incidence x-ray diffraction (GIXRD) of the film is recorded on a Rigaku Smartlab high-resolution X-ray diffractometer (Cu-Kα source with λ = 1.54 Å) at 0.6° grazing angle using a scan rate of 1°/min with 0.01° step size in the 2θ range 15°-70°. The thickness of the film and SiO$_x$ interface layer are determined using x-ray reflectivity (XRR) measurement. The x-ray photoelectron spectroscopy (XPS) is also performed for the elemental study. The top blanket W-metal is etched out before GIXRD and XPS measurements. The Keysight B1500A semiconductor device parameter analyzer is used to measure the DC current voltage (I-V) and small signal capacitance-voltage (C-V) characteristics. Finally, the polarization-voltage (P-V) hysteresis loop, Positive-Up-Negative-Down (PUND) and endurance measurements are carried out in a Precision LC-II ferroelectric tester of Radiant Technologies.

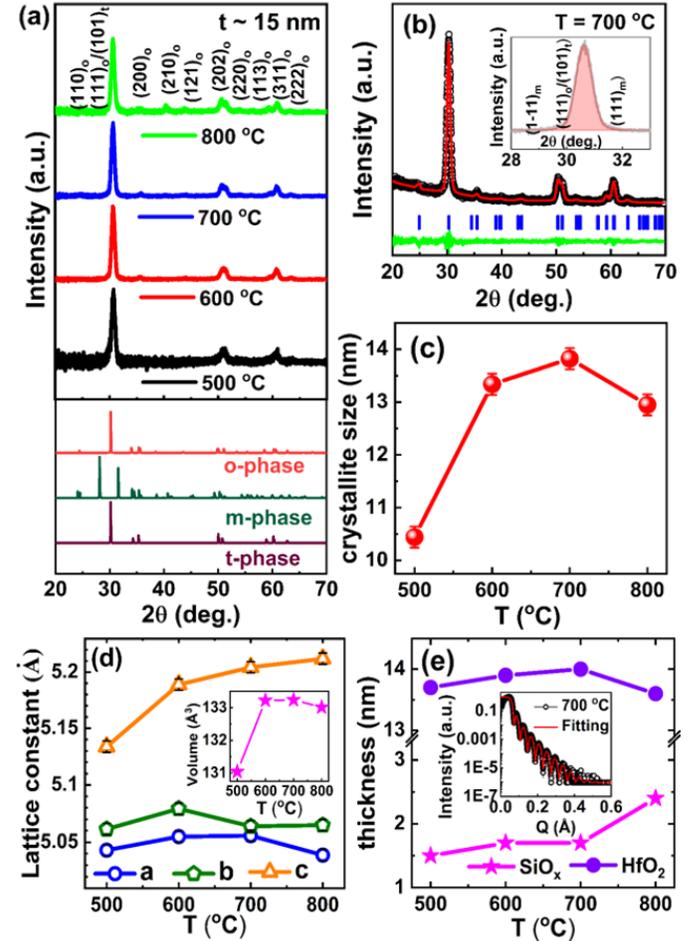

**Figure 2:** Annealing temperature (T) effect on (a) GIXRD spectrum of HfO$_2$ films which shows the main peak at ~30.4° of all samples are indexed with orthorhombic peak of HfO$_2$ which is basically responsible for ferroelectricity in HfO$_2$. (b) Lebail fitting and deconvoluted GIXRD spectra of T = 700 °C (for example), (c) Crystallite size (D$_{cs}$) increases with T. D$_{cs}$ is calculated using Scherrer formula, (d) Lattice constants (a, b, and c) calculated using Le-bail fitting of GIXRD pattern considering o-phase where c-axis increases with temperature and saturates at higher T. Inset figure shows volume expansion corresponds to strain increment in the film. (e) HfO$_2$ film thickness and interfacial layer (SiO$_x$) from XRR fitting (in-set figure). At higher temperature, interfacial reaction increases. GIXRD measurement was performed on annealed samples after etching top electrode (W).

## III. RESULTS AND DISCUSSION

### A. Effect of annealing temperature

In first part of the experiment, we investigate the effect of annealing temperature (T) on the structural and ferroelectric properties of the HfO$_2$ film. The GIXRD and XRR measurements are carried out to examine the phase purity, crystal structure and thickness of the film (fig. 2). Fig. 2(a) shows the GIXRD pattern of HfO2 film with standard HfO2 polymorph database. It reveals that all the peaks are well matched with o-/t-phases. The deconvolution of peak at 2θ ~ 30.4°, shown in fig.-2(b) indicates the suppression of m-phase and enhancement of o-phase due to induced biaxial tensile strain on the film by post metallization annealing (PMA). Crystallite size ($D_{cs}$) for all the samples is calculated using the Debye–Scherrer formula ($D_{cs} = \frac{K.\lambda}{\beta_{film} cos\theta}$, where $K = 0.94$ Scherrer's constant, $\beta_{film} = \sqrt{\beta_{total}^2 - \beta_{Si}^2}$ is full-width-half-maxima (FWHM) of the film, $\beta_{Si}$ is FWHM of standard Si sample) [21]. In fig. 2(c), the variation of crystallite size is plotted with T which depicts the steep increase of $D_{cs}$ from 500 °C by 40% to plateau in the 600-800 °C. This significant increment of $D_{cs}$ can be attributed to the increased crystallization of the film with increasing T [22]. The plateau of $D_{cs}$ in 600-800 °C indicates completion of crystallization without any further grain growth. However, $D_{cs}$ decreases slightly (by ~7%) at 800 °C compared to 700 °C may be due to thicker $SiO_x$ formation at the Si/HfO2 interface. The lattice constants (a, b and c) are calculated using Le-bail fitting of GIXRD pattern considering o-phase ($Pca2_1$) [23]. Figure-2(d) shows that 'c' increases from 5.13(±0.01) to 5.22(±0.01) Å with increasing T, whereas 'a' and 'b' are unequal and show marginal variation. The c-parameter increment with T matches well with the polarization increment with T shown later. The calculated lattice parameter values are in good agreement with the previously reported lattice parameters of orthorhombic phase of HfO2 [24]–[29]. Table-I tabulates the calculated lattice parameters of orthorhombic HfO2 with earlier reported state of the art results. Also, the volume expansion of the orthorhombic unit cell (inset of fig. 2(d)) depicts similar behavior as crystallite size change with T shown in fig. 2(c). The thickness variations of HfO2 and interfacial layer ($SiO_x$) with T are plotted in Fig. 2(e) which are extracted from the XRR data fitting (inset fig. 2(e)). HfO2 thickness remains almost constant. It slightly increases from ~13.7 (±0.2) nm to ~14.0 (±0.2) nm for 500-700 °C and falls to 13.6 (±0.2) nm at 800 °C correlates well with thicker $SiO_x$ layer (~ 2.5(±0.1) nm) compared to ~1.6 nm at lower temperatures, indicating temperature accelerated interfacial reaction.

TABLE I
COMPARISON OF LATTICE CONSTANTS *a, b,* AND *c* AND UNIT CELL VOLUME *V* OF ORTHORHOMBIC UNIT CELL OF HfO2 WITH STATE OF THE ART LITERATURE RESULTS

| Orthorhombic Unit cell | | Lattice parameters in Å | | | Unit cell volume Å³ |
|---|---|---|---|---|---|
| Ref. | Space Group | a | b | c | V |
| [25] | $Pca2_1$ | 5.04 | 5.06 | 5.23 | 133.377 |
| [26] | $Pca2_1$ | 5.063 | 5.093 | 5.291 | 136.432 |
| [27] | $Pca2_1$ | 5.007 | 5.055 | 5.224 | 132.221 |
| [28] | Ortho I HfO2 | 5.007 | 5.058 | 5.227 | 132.375 |
| [29] | $Pca2_1$ | 5.003 | 5.059 | 5.228 | 132.321 |
| This work | **600 °C ($Pca2_1$)** | **5.055** | **5.081** | **5.19** | **133.302** |
| | **700 °C ($Pca2_1$)** | **5.055** | **5.065** | **5.22** | **133.65** |

**Bold rows are experimental data; others are theoretical**

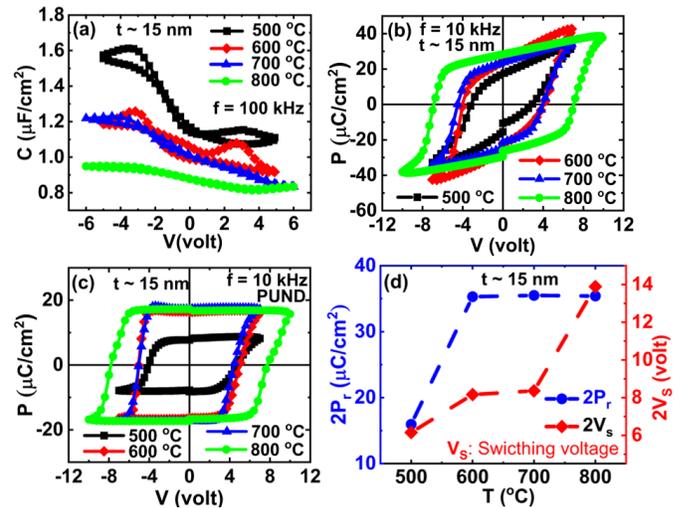

Figure 3: Influence of Annealing temperature (T) on (a) Capacitance vs. voltage characteristics measured at 100 kHz of small signal frequency with 30 mV amplitude, (b) Polarization vs. voltage (P-V) characteristics, Polarization is measured at 10 kHz with triangular pulse, (c) PUND P-V hysteresis characteristics shows increment of remnant polarization ($2P_r = +P_r + |-P_r|$) from 15.95 μC/cm² to 35.5 μC/cm² and then saturates at higher T and (d) $2P_r$ and switching voltage ($2V_S$) for HfO2 (~15 nm) films with respect to T reveals higher $2P_r$ with lower $2V_S$ are resulted from 600 and 700 °C samples whereas 800 °C sample shows very high switching voltage which is may be due to thicker $SiO_x$ formation. Five devices were measured from every sample and averaged them.

In Fig. 3(a), butterfly loops observed in the C-V characteristics are the signature of ferroelectric switching where the polarization flips at peak positions which are superimposed with a typical metal oxide semiconductor (MOS) C-V where ferroelectric capacitance ($C_{HfO2}$) and interfacial layer capacitance ($C_{SiOx}$) are in series combination. As with increase in anneal temperature (T), low-k interfacial oxide thickness ($SiO_x$) increases and hence drop across it also increases which causes the reduced voltage drop across HfO2 thus less polarization switching happens and due to this butterfly peaks diminish at higher T. The measured P-V for all samples at different T are plotted in Fig. 3(b) which confirms ferroelectricity in these intrinsic HfO2 films. PUND measurement is also carried out to extract the pure ferroelectric polarization arising from the domain switching. A well-defined PUND hysteresis loops of undoped-HfO2 annealed at different temperatures are shown in Fig. 3(c). Fig. 3(d) depicts that the remnant polarization ($2P_r = +P_r + |-P_r|$) is ~15.95 μC/cm² for 500 °C sample and it increases to ~35.5 μC/cm² and remains almost constant for 600-800 °C**.** The polarization observed in this study is higher in comparison to earlier reported undoped-HfO2 films [14], [15], [17], [21], [30]–[33]. The volume expansion of the orthorhombic unit cells and crystallite size increment observed using XRD correlate and explain the enhancement of polarization in the HfO2 film with T. The switching voltage ($V_S$) for 600-700 °C samples are similar ($2V_S$ ~ 8 V) shown in fig. 3(d), however for 800 °C sample, the $2V_S$ is very high ~14 V among all four temperatures because of the thicker SiO$_x$ formation at the Si/HfO$_2$ interface as indicated from XRR measurements.

## B. Effect of film thickness

In the second part, we examine the influence of thickness from 5 nm to 29 nm on the ferroelectric properties of the films annealed at optimal temperature window. Fig. 4(a) and (b) show the GIXRD pattern for various thicknesses annealed at 600 and 700 °C, respectively. At both the temperatures, 5 to 20 nm films show clear orthorhombic peaks at ~30.4°. In contrary, the 29 nm films reveal presence of m-phase (star marked) in addition to o-peaks which may be due to stress relaxation at higher thickness as observed earlier [14] [34]. The m-phase percentage is calculated as: %m − phase = $\frac{I_m}{I_m + I_o}$, (where $I_m$ is intensity of (1-11)$_m$ peak and $I_o$ is intensity of (111)$_o$ peak) [35]. The calculated percentage of m-phase present in 29 nm samples annealed at 600 and 700 °C are ~ 6.6 (±0.4) % and ~13 (±0.6) %, respectively, which indicates a 2× increase of m-phase with temperature. In addition, the peak broadening is seen in 5 nm films for both temperatures corresponding to smaller D$_{cs}$. Fig. 4(c) shows that the D$_{cs}$ increases from ~4 (±0.2) nm to ~17 (±0.2) nm with increasing thickness whereas, there is saturation (600 °C) or even a little fall off (700 °C) at 29 nm. Thus, we observe that D$_{cs}$ of the o-phase scales with thickness in the range 5-20 nm. This correlation indicates that the crystallization occurs across the entire film thickness. The D$_{cs}$ scaling with thickness of the o-phase saturates (or even reduces) beyond 20 nm suggesting the m-phase formation at the expense of o-phase crystallite growth. Thus, we discover that 20 nm is a critical thickness for pure o-phase to m-phase transformation. Fig. 4(d) clearly shows the lattice constants a, b, and c are unequal and increase from 10 nm to 15 nm sample (indicating increased strain) and saturate till 20 nm for both the annealing temperatures (T) and then falls for 29 nm samples (reduction in strain). The reduction in lattice constants at higher thickness (29 nm) is due the m-phase increment as seen in XRD. It is also seen that lattice constant 'c' is maximum for 15 and 20 nm films at both T indicating the larger tensile strain over others. We will later see that it correlates well with measured 2P$_r$ trends. Further, XPS is performed to determine the valance state of the elements and chemical composition of the films.

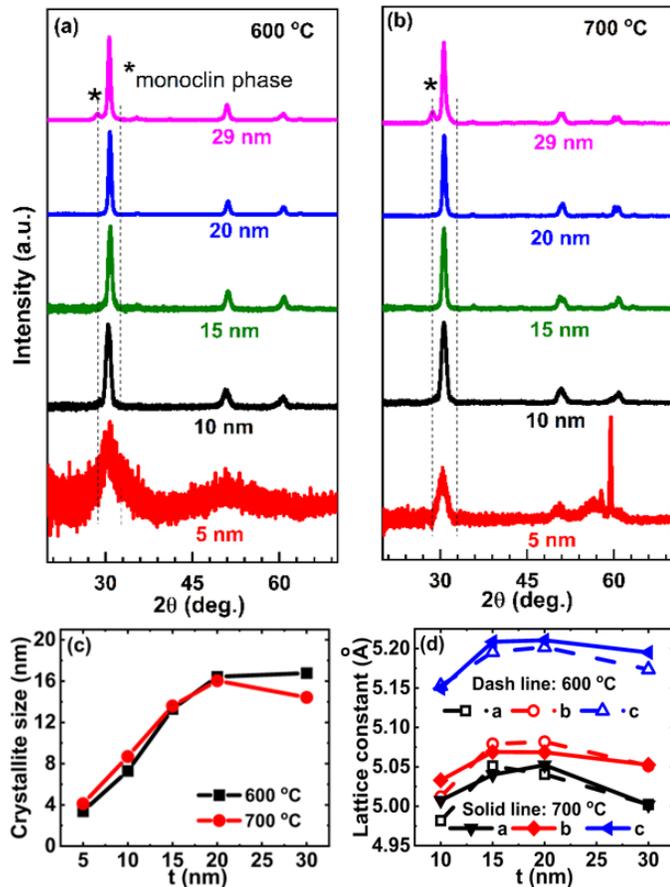

**Figure 4:** (a) Film thickness effect on GIXRD for (a) T = 600 °C and (b) 700 °C annealed HfO$_2$ films which shows peaks at ~30.4°. All samples are indexed with orthorhombic peak of HfO$_2$ with only 29 nm films at both temperatures show monoclinic peaks as well which contributes ~ 6.8 % and ~ 13 % of m-phase (c) Crystallite size (D) increases with film thickness up to 20 nm for both the temperatures whereas D falls a little bit off for 700 °C sample and this might be the reason for higher m-phase present in this sample. (d) Lattice constants (a, b, and c) slightly increase with film thickness up to 20 nm for both the temperatures and then all of them start falling. Lattice constants are calculated using Le-bail fitting of GIXRD pattern considering o-phase.

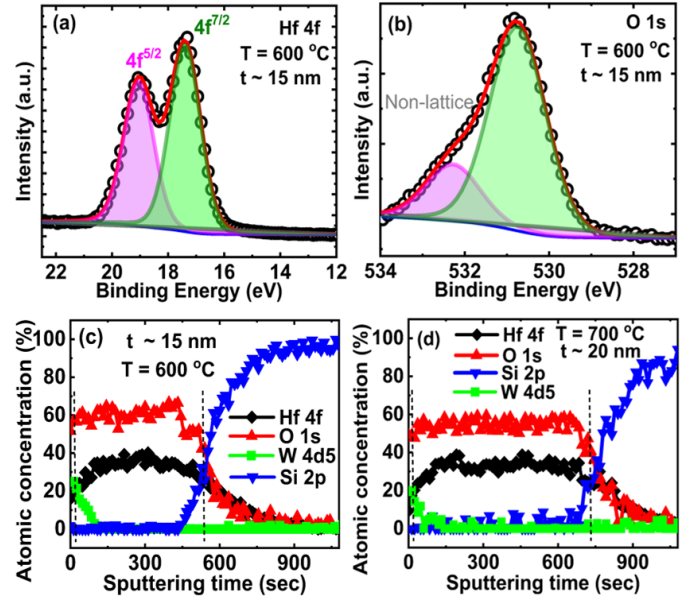

**Figure 5:** The high-resolution XPS fitting results of Hf-4f and O-1s peaks for 15 nm film annealed at 600 °C are shown in (a) and (b), respectively, with experiment data (black points), fitting result (red sol) and (c) and (d) are the depth profile plots of 15 and 20 nm films annealed at 600 and 700 °C respectively. XPS measurement was done on annealed samples after etching top electrode (W).

The high-resolution surface and depth XPS analysis of Hf-4f and O-1s peaks for the 15 and 20 nm films annealed at 600 and 700 °C are conducted.

However, there is no significant difference is observed in the XPS data of those films. The representative XPS spectra of 15 nm film annealed at 600 °C for Hf-4f and O-1s is shown in Fig. 5 (a) and (b), respectively. The Hf-4f doublet peaks with binding energy (BE) located at 17.32 (±0.02) eV and 19.07 (±0.3) eV (with a spin–orbit splitting of 1.75 eV) are assigned to Hf-4f$^{7/2}$ and Hf-4f$^{5/2}$, respectively, indicating Hf-O bond formation in HfO$_2$ film [36]. In fig. 5(b), the O-1s peak at ~530.82 (±0.02) eV is associated with the lattice oxygen, revealing the formation of Hf-O bond, and the peak at higher BE ~532.32 (±0.03) eV corresponds to the non-lattice oxygen associated to the overlap of surface hydroxyl (OH) groups and contaminants containing oxygen [37][38]. The high percentage of the non-lattice oxygen from the O-1s spectra indicates more defects (OH or O$^{\delta-}$) present in the HfO$_2$ film [38]. Figure 5 (c) and (d) depict the XPS depth profile of the 15 and 20 nm films annealed at 600 and 700 °C respectively with a sputtering step of 15 s. They clearly show the formation of HfO$_2$ along with residual WO$_x$ from top interface and SiO$_x$ at the bottom interface in both the samples.

variation in Fig. 6(c) highly correlates to lattice constants (volume) variation with film thickness (shown in Fig. 4(d)). The temperature insensitivity of 2P$_r$ in the 600-700 °C range for films below critical thickness of 20 nm reveals complete o-phase formation. However, above the critical thickness 700 °C (cf. 600 °C) annealing shows a higher reduction in 2P$_r$ indicating a higher o→m-phase transition.

Further, endurance test is performed for 15 nm films annealed at 600 and 700 °C as plotted in Fig.-6(d) which reveals a wake-up free high endurance of ~10$^8$ cycles for both the samples with 2P$_r$ window of ~17 µC/cm$^2$ and none of the devices breakdown even after 10$^8$ cycles. Wakeup is the gradual increment of remnant polarization due to the transformation of t-phase to o-phase and oxygen defects distribution during field cycling [39] which is absent in our devices and this clearly indicates the complete phase transformation to o-phase during annealing itself.

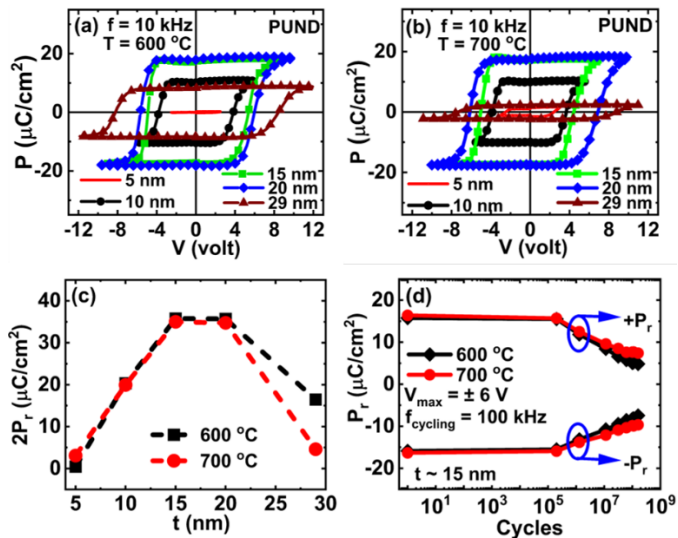

**Figure 6**: The effect of film thickness on remnant polarization and switching voltage are seen in the PUND hysteresis loops of the samples annealed at (a) 600 °C and (b) 700 °C respectively. Fig. (c) shows the variation of remnant polarization (2P$_r$) calculated from fig. (a) and (b). (d) Endurance characteristics with triangular pulse of 100 kHz cycling frequency and V$_{max}$ = ±6 V while reading is done by PUND measurement with same amplitude of triangular pulse and 10 kHz frequency.

Figure 6(a) and 6(b) depict the PUND measurements of the various thicknesses of HfO$_2$ annealed at 600 and 700 °C respectively. At both the annealing temperatures, switching voltage increases as film thickness increases and remnant polarization 2P$_r$ steeply increase from 3 to 36 µC/cm$^2$ with increase of thickness from 5 nm to 15 nm samples and then nearly saturates to ~35.5 µC/cm$^2$ for 20 nm samples summarized in fig. 6(c). However, the 2P$_r$ value falls drastically for 29 nm samples for both temperatures due to the formation of m-phase as shown in XRD. Also, 2P$_r$

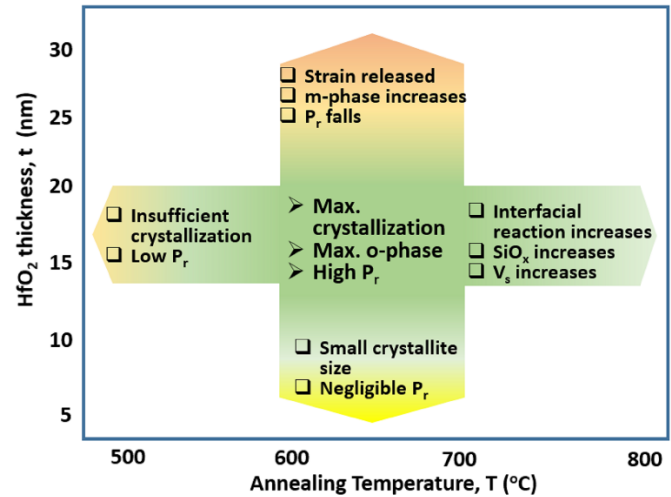

**Figure 7**: Summary plot of the whole experiment outcome revealing temperature-thickness window (T = 600-700 °C and thickness = 15-20 nm) with well crystallized o-phase and enhance polarization.

Finally, Fig. 7 summarizes the whole experiment outcome. First, the temperature rise leads to improved crystallization and ferroelectricity at the cost of interlayer SiO$_x$ thickness (beyond 700 °C) reflected in the higher switching voltage. Second, with increase in thickness, crystallite size and lattice constants (volume) increase till a critical thickness of 20 nm which correlates well with higher 2P$_r$. Beyond the critical thickness, crystallite size scaling with thickness saturates and unwanted monoclinic phase starts appearing which indicates a 2× increase in m-phase with temperature for 29 nm sample at T = 700 °C.

Comparing with state-of-the-art results (Table-II), the proposed undoped-HfO$_2$ film shows higher remnant polarization and endurance at lower processing temperatures (200 °C lower). The Table-II reveals that the proposed films show higher 2P$_r$ and endurance than the reported MFIS results. Benchmarking Table-II also shows that Si BE and TiN TE produce highest 2P$_r$ ~38 µC/cm$^2$

TABLE II
BENCHMARKING WITH STATE OF THE ART WORK

| Ref. No. | Device Stack | Dep. Process | t in nm | $T_d$ in °C | Anneal condition | $2P_r$ (μC/cm²) | Endurance |
|---|---|---|---|---|---|---|---|
| [14] | MFM | ALD | 6 | 300 | 650 °C, 30s, $N_2$ | 20 | $1.5 \times 10^5$ |
| [15] | MFM | ALD | 9 | 220 | 650 °C, 30s, $N_2$ | 20.8 | $10^8$ |
| [30] | MFIS | ALD | 6 | 275 | 800 °C, 30s, $N_2$ | 38 | NA |
| [31] | MFM | ALD | 10 | 220 | 650 °C, 60s, $N_2$ | 16 | NA |
| [32] | MFM | ALD | 8.3 | 260 | 650 °C, $N_2$ | 27 | NA |
| [40] | MFIS | ALD | 6 | 275 | 800 °C, 30s, $N_2$ | 37.5 | NA |
| [33] | MFM | PLD | 7.4 | 780 | No anneal | 29.4 | NA |
| [17] | MFM | Sputter | 10 | RT | 800 °C, 20s, $N_2$ | 24 | NA |
| [19] | MFM | Sputter | 20 | RT | 800 °C, 20s, $N_2$ | 20 | $10^8$ |
| **This work** | **MFIS** | **Sputter** | **15** | **RT** | **600 °C, 30s, $N_2$** | **35.5** | **$10^8$** |

MFM: Metal/HfO₂/Metal, MFIS: Metal/HfO₂/SiO$_x$/Si, RT: Room temperature, $T_d$: Deposition temperature, t: film thickness, NA: Not available.

[30], [40] for ALD-HfO₂ with 800 °C annealing temperature. However, we show similar $2P_r$ ~36 μC/cm² with room temperature deposition and lower annealing temperature (600 °C) in sputtered-HfO₂ film. Hence, Si BE is the important contributor due to its low thermal expansion coefficient (Si: TEC ~ 2.6 μm.m$^{-1}$.K$^{-1}$). The W TE with low TEC ~ 4.5 μm.m$^{-1}$.K$^{-1}$ also leads to out of plane confinement which induces high biaxial tensile strain on HfO₂ film. This confinement increases the o-phase formation during annealing and thereby enhancing ferroelectric polarization [39], [41], [42]. Further, the optimal films show wake-up free robust endurance of $10^8$ cycles, which is demonstrated for the first time for such high $2P_r$ films [30], [40]. Given, the high quality of ferroelectric films identified in the optimal temperature-thickness window, the thickness and temperature studies become highly relevant.

## IV. CONCLUSION

In summary, we have investigated the ferroelectric properties of intrinsic HfO₂ directly deposited on Si substrate by RF sputtering and experimentally demonstrated the effect of the annealing temperature and film thickness. We have demonstrated a temperature-thickness window (600-700°C and 15-20 nm) which shows a high remnant polarization $2P_r$ of ~35.5 μC/cm² and wake-up free robust endurance up to ~$10^8$ cycles. The polarization of the films is greatly correlated with the physical properties viz, phase, lattice constant, crystallite size and interface reaction. Our films show state-of-the-art polarization, which enhances the relevance of the thickness and temperature-based studies. Further optimization of the process parameters and device stack engineering can help in achieving ferroelectric properties comparable to doped HfO₂. Ferroelectric undoped HfO₂ film grown by sputtering on Si can be used to fabricate and realize the energy efficient FeFET and FTJ based non-volatile crossbar array for high density storage applications.

ACKNOWLEDGMENT

The authors are grateful to Shashank Inge and Vivek Saraswat of IIT Bombay for the fruitful discussion and suggestions. The authors are also grateful to IRCC, IIT Bombay for the XRD measurement facility.